\documentstyle[12pt]{article}

\begin{document}


\title{ Comments on QCD Corrections to  $b \to s \gamma$ Decay
\thanks{ Partly supported by the National
Natural Science Foundation of China
 and the Doctoral Program Foundation of Institution of Higher
Education. }}

\author{  Cai-Dian L\"{u}$^{a}$\thanks{E-mail
Address: lucd@itp.ac.cn} Jing-Liang Hu$^{b}$ and Chong-Shou Gao$^{a,b}$\\
$a$ Institute of Theoretical Physics, Academia Sinica,\\
P.O.Box 2735, Beijing 100080, China\\
$b$ Physics Department, Peking University,
Beijing 100871, China\thanks{Mailing address} }

 \date{\today}
\maketitle

\begin{abstract}
We find some errors in previous calculation of leading log QCD
 corrections to $b \to s\gamma $ decay, which include corrections
from $m_{top}$ to $M_W$ in addition to corrections  from $M_{W}$ to $m_b$.
The inclusive decay rate is found to be enhanced more than
previous calculations. At $m_t=170$GeV, the
running from  $m_{top}$ to $M_W$ results in 13\% enhancement, and for
$m_t=250$GeV, 16\% is found.
\end{abstract}


\newpage

It is well known that the process $b \to s\gamma $ is extremely sensitive
to new physics beyond the Standard Model.
In 1993, the CLEO collaboration placed an  upper limit
on the inclusive $b \rightarrow s \gamma$ decay
$B(b \rightarrow s \gamma)<5.4 \times 10^{-4}$ at 95\%
C.L.\cite{cleo2}.
This has inspired a large number of papers\cite{Hew}.
It has been argued that this experiment provides more
information about restrictions on the Standard Model, 2-Higgs
doublet model, Supersymmetry, Technicolor and etc.
Recently CLEO has measured the inclusive branching ratio to be\cite{cleo}
\begin{equation}
Br( b \to s \gamma )=(2.32\pm 0.51 \pm 0.29 \pm 0.32) \times 10 ^{-4}.
\end{equation}
Corresponding to 95\% confidence level, the range is
$1 \times 10^{-4} < Br(b \to s \gamma )<4\times 10 ^{-4}$. More stringent
constraints are obtained by experiments, so more accurate
theoretical calculation of this decay rate is needed.

The radiative $b$ quark decay has already been calculated
 in many papers\cite{Springer,Mis,Cho}.
It is found to be strongly QCD-enhanced.
In other words, the strong interaction plays an important
role in this decay. However, there are still some uncertainties in
these  papers. Most papers\cite{Springer,Mis}
do not include the QCD running from $m_{top}$
to $M_W$. Since the top quark is found to be heavier than W boson(
$m_{top} = 174\pm 10 ^{+13}_{-12}$ GeV \cite{D0} ),
it needs a detailed calculation of this effect.
Ref.\cite{Cho} does include this running, however there
are some errors in calculation of anomalous dimensions which may lead
to some changes in final result.

In our present paper, by using effective field theory
 formalism\cite{Georgi},
we recalculate the $b \to s \gamma$ decay in Minimal Standard Model.
We first integrate out the top quark, generating an effective
five-quark theory. By using the
renormalization group equation, we run the effective field theory
down to the W-scale, including QCD corrections from
$m_{top}$ to $M_W$, and correct some errors in ref.\cite{Cho}.
Then the weak bosons are removed.
Untruncated anomalous dimensions of QCD running from
$M_W$ to $m_b$ are used.
Finally we calculate the rate of radiative $b$ decay.

The effective Hamiltonian is written as
\begin{equation}
{\cal H}_{eff}=2 \sqrt{2} G_F V_{tb}V_{ts}^*\displaystyle \sum _i
C_i(\mu)O_i(\mu). \label{eff}
\end{equation}
The operators $O_i$ make a complete basis of dimension-6 operators:
\begin{eqnarray}
O_{LR}^1  & =  &  -\frac{1}{16\pi^2} m_b \overline{s}_L D^2 b_R,
\nonumber\\
O_{LR}^2  &  =  &  \mu^{\epsilon/2} \frac{g_3}{16\pi^2}
	m_b \overline{s}_L \sigma^{\mu\nu} X^a b_R G_{\mu\nu}^a,
\nonumber\\
O_{LR}^3  &  =  &  \mu^{\epsilon/2} \frac{e Q_b}{16\pi^2}
	m_b \overline{s}_L \sigma^{\mu\nu} b_R	F_{\mu\nu},
\nonumber\\
Q_{LR}  &  =  &  \mu^{\epsilon} g_3^2 m_b
	\phi_{+}\phi_{-} \overline{s}_L b_R,
\nonumber\\
P_L^{1,A}  &  =  &  -\frac{i}{16\pi^2}  \overline{s}_L
  T_{\mu\nu\sigma}^A D^{\mu} D^{\nu} D^{\sigma} b_L,\nonumber\\
P_L^2 & = & \mu^{\epsilon/2} \frac{e Q_b}{16\pi^2}  \overline{s}_L
	\gamma^{\mu} b_L \partial^{\nu} F_{\mu\nu},\nonumber\\
P_L^3  &  =  &  \mu^{\epsilon/2} \frac{e Q_b}{16\pi^2}  F_{\mu\nu}
	\overline{s}_L 	\gamma^{\mu} D^{\nu}b_L,\nonumber\\
P_L^4  &  =  &  i \mu^{\epsilon/2} \frac{e Q_b}{16\pi^2}
	\tilde{F}_{\mu\nu}
	\overline{s}_L \gamma^{\mu} \gamma^5 D^{\nu} b_L,\nonumber\\
R_L^1 &  =  &  i \mu^{\epsilon} g_3^2 \phi_{+}\phi_{-} \overline{s}_L
	\not \!\! D b_L,\nonumber\\
R_L^2  &  =  &  i \mu^{\epsilon} g_3^2(D^{\sigma} \phi_+) \phi_{-}
	\overline{s}_L\gamma_{\sigma} b_L,
\nonumber\\
R_L^3  &  =  &  i \mu^{\epsilon} g_3^2 \phi_{+}(D^{\sigma} \phi_{-})
	\overline{s}_L\gamma_{\sigma} b_L.
\end{eqnarray}
The coefficients $C_i(\mu =m_t )$ are calculated from matching diagrams,
and agree with ref.\cite{Cho}.

After evaluating the loop diagrams, we find that the
weak mixing of operators agrees with ref.\cite{Cho}.
While the QCD anomalous dimensions for each of the operators
in our basis are
\begin{equation}
 \begin{array}{c}
     \begin{array}{ccccccccccccc}
	& && O_{LR}^1 & O_{LR}^2& O_{LR}^3& P_L^{1,1}& P_L^{1,2}&
	  P_L^{1,3}& P_L^{1,4}& P_L^{2}& P_L^{3}& P_L^{4}
     \end{array}\\
     \begin{array}{r}
  O_{LR}^1\\ O_{LR}^2\\ O_{LR}^3\\ P_L^{1,1}\\ \gamma=\; P_L^{1,2}\\
	  P_L^{1,3}\\ P_L^{1,4}\\ P_L^{2}\\ P_L^{3}\\ P_L^{4}
     \end{array}\left(\begin{array}{ccccccccccccccc}
	  \frac{20}{3} && 1 && -2 & 0 & 0 & 0 & 0 && 0 & 0 && 0 \\
 -8 && \frac{2}{3} && \frac{4}{3} & 0 & 0 & 0 & 0 && 0 & 0 && 0 \\
	 0 && 0 && \frac{16}{3} & 0 & 0 & 0 & 0 && 0 & 0 && 0 \\
	 6 && 2 && -1 & \frac{2}{3} & 2 & -2 & -2 && 0 & 0 && 0 \\
	 4 && \frac{3}{2} && 0 & -\frac{113}{36} & \frac{137}{18}
	 & -\frac{113}{36} &-\frac{4}{3} &&\frac{9}{4} & 0 && 0  \\
	 2 && 1 && 1 & -2 & 2 & \frac{2}{3} & -2 && 0 & 0 && 0 \\
	 0 && \frac{1}{2} && 2 & -\frac{113}{36} & \frac{89}{18}
	  & -\frac{113}{36} &
	 \frac{4}{3} && \frac{9}{4} & 0 && 0 \\
	 0 && 0 && 0 & 0 & 0 & 0 & 0 && 0 & 0 && 0 \\
	 0 && 0 && -\frac{4}{3} & 0 & 0 & 0 & 0 && 0 & 0 && 0 \\
	 0 && 0 && -\frac{4}{3} & 0 & 0 & 0 & 0 && 0 & 0 && 0
	\end{array}\right) \displaystyle{ \frac{g_3^2}{8\pi^2} },
\end{array}
\label{anom1}
\end{equation}
\begin{equation}
\begin{array}{cccc}
\gamma= & \begin{array}{c}
	\\ Q_{LR}\\ R_L^1\\ R_L^2\\ R_L^3\\
          \end{array}

	& \begin{array}{c}
	      \begin{array}{cccc}Q_{LR} & R_L^1 & R_L^2 & R_L^3
 		     \end{array}
	    \\
	      \left( \begin{array}{cccccc}
 \frac{23}{3} && 0  & 0  &&  0\\
 0  && \frac{23}{3} & 0  &&  0\\
 0  &&  0 & \frac{23}{3} && 0\\
 0  &&  0 & 0 && \frac{23}{3}\\
	      \end{array} \right)
	  \end{array}
	& \displaystyle{ \frac{g_3^2}{8 \pi^2} }.

\end{array}\label{anom2}
\end{equation}
Comparing with ref.\cite{Cho}, there are some differences in the
anomalous dimension matrix, which may lie in omiting a symmetric factor
 of 1/2 in ref.\cite{Cho} in calculating Feynman diagram like Fig.1.
And some changes may due to miscalculation.

 After these changes, the whole matrix can be easily diagonalized,
and gives all real eigenvalues while that in ref.\cite{Cho} can not.
Inserting anomalous dimension (\ref{anom1})(\ref{anom2})
to the renormalization group equation satisfied by
the coefficient functions $C_i(\mu)$
\begin{equation}
\mu \frac{d}{d\mu} C_i(\mu)=\displaystyle\sum_{j}(\gamma^{\tau})_{
ij}C_j(\mu).\label{ren}
\end{equation}
we can have the coefficients of operators at
$\mu=M_W$. And some of these operators
change a lot from ref.\cite{Cho} due to our improvements.


In order to continue running the basis operator coefficients down to
lower scales, one must integrate out the weak gauge bosons and
would-be Goldstone bosons at $\mu=M_W$ scale. This leads to the
 well-known six four-quark operators\cite{Springer,Mis}.
The other remaining two-quark operators can be reduced
by using equations of motion(EOM) to the gluon and photon magnetic moment
operators $O_{LR}^2$ and $O_{LR}^3$.

To be comparable with previous results, we rewrite our operators
$O_{LR}^3$, $O_{LR}^2$ as $O_7$, $O_8$ like ref.\cite{Mis},
\begin{eqnarray}
O_7&=&(e/16\pi^2) m_b \overline{s}_L \sigma^{\mu\nu}
	    b_{R} F_{\mu\nu},\nonumber\\
O_8&=&(g/16\pi^2) m_b \overline{s}_{L} \sigma^{\mu\nu}
	    T^a b_{R} G_{\mu\nu}^a.
\end{eqnarray}

For completeness, we give the explicit expressions of
the coefficient of operator $O_8$ and
$O_7$ at $\mu=M_W^-$,
\nopagebreak[1]
\begin{eqnarray}
C_{O_8}(M_W^-) &= & \left( \frac{\alpha _s (m_t)} {\alpha _s (M_W)}
	\right) ^{ \frac{14}{23} } \left\{ \frac{1}{2}C_{O_{LR}^1}(m_t)
-C_{O_{LR}^2}(m_t) +\frac{1}{2}C_{P_L^{1,1}}(m_t) \right.\nonumber\\
&&	\;\;\;\;\;\;\;\;\;\;\;\;\;\;\;\;\;\;
	\left.+\frac{1}{4}C_{P_L^{1,2}}(m_t)
	-\frac{1}{4}C_{P_L^{1,4}}(m_t)\right\}
	-\frac{1}{3} ,\label{c2}
\end{eqnarray}
\begin{eqnarray}
&\displaystyle{
C_{O_7}(M_W^-) = \frac{1}{3}\left( \frac{\alpha _s (m_t)} {\alpha _s (M_W)}
	\right) ^{ \frac{16}{23} } \left\{ C_{O_{LR}^3}(m_t)
	+8 C_{O_{LR}^2}(m_t) \left[1-\left( \frac{\alpha _s (M_W)}
{\alpha _s (m_t)} \right) ^{ \frac{2}{23} } \right]\right.}&\nonumber\\
&\displaystyle{	+\left[-\frac{9}{2} C_{O_{LR}^1}(m_t)
	-\frac{9}{2}C_{P_L^{1,1}}(m_t)
-\frac{9}{4}C_{P_L^{1,2}}(m_t) +\frac{9}{4}C_{P_L^{1,4}}(m_t)\right]
\left[1- \frac{8}{9} \left( \frac{\alpha _s (M_W)}
{\alpha _s (m_t)} \right) ^{ \frac{2}{23} } \right] }&\nonumber\\
&\displaystyle{	\;\;\;\left.-\frac{1}{4}C_{P_L^4}(m_t)
+\frac{9}{23} 16\pi^2 C_{W_L^1}(m_t) \left[1- \frac{\alpha _s (m_t)}
	{\alpha _s (M_W)} \right] \right \}
	-\frac{23}{36} }. &  \label{c3}
\end{eqnarray}
They are expressed by coefficients of operators
at $\mu=m_t$ and QCD coupling $\alpha_s$.
So it is convenient to utilize these formula.

The obvious differences from QCD correction to $C_7(M_W)$ and
$C_8(M_W)$ can easily be seen from Fig.2.
In comparison to ref.\cite{Cho}, the enhancement of
coefficient of operator $O_7$ is almost  the same size,
but the values for $O_8$ are quite different. Here the effect to
$O_8$  is an enhancement rather than a suppression as in
ref.\cite{Cho}. These changes come from the corrections
of anomalous dimensions described earlier. Since $C_7(M_W)$
and $C_8(M_W)$ are both the input of the following QCD
running from $M_W$ to $m_b$, it is expected to change the final result.

The running of the coefficients of operators from $\mu=M_W$ to
$\mu=m_b$ was well described in ref.\cite{Mis}. After this running
 we have the coefficients of operators at $\mu=m_b$ scale.
Here we use $M_W=80.22$GeV, $m_b=4.9$GeV. Both $C_7(m_b)$ and
$C_8(m_b)$ are enhanced in comparison to values obtained by Misiak
where the QCD corrections from $M_W$ to $m_t$ are neglected\cite{Mis}.


The leading order $b \rightarrow s\gamma$ matrix element of $H_{eff}$
is given by the sum of operators $O_5$, $O_6$ and $O_7$, this
disagrees with ref.\cite{Cho}, but agrees with Misiak\cite{Mis}.
The sought amplitude will be proportional to the squared
 modulus of
\begin{equation}
	C_7^{eff}(m_b)=C_7(m_b)+Q_d\;[C_5(m_b)+3 C_6(m_b)] \label{C7}
\end{equation}
instead of $|C_7(M_b)|^2$ itself.

Following ref.\cite{Springer,Mis},
applying eqs.(\ref{C7}), one finds
\begin{equation}
\frac{BR(\overline{B} \rightarrow X_s \gamma)}{BR(\overline{B}
\rightarrow X_c e \overline{\nu})} \simeq \frac{6 \alpha_{QED}}{\pi
 g (m_c/m_b)}
|C_7^{eff}(m_b)|^2 \left(1-\frac{2 \alpha_{s}(m_b)}{3 \pi} f(m_c/m_b)
\right)^
{-1},
\end{equation}
where $g(m_c/m_b)\simeq 0.45$ and $f(m_c/m_b)\simeq 2.4$ corresponding
to the phase space
factor and the one-loop QCD correction to the semileptonic decay,
respectively\cite{Cabi}. The electromagnetic fine structure constant
evaluated at the $b$ quark scale takes value as $\alpha_{QED}(m_b)=
1/132.7$. The results are summarized in Fig.3 as functions of
 the top quark mass. The  QCD-uncorrected values are also shown.
In this figure, one can easily see that, at $m_t=170$GeV, it results
in 13\% enhancement from Misiak's result\cite{Mis},
and for $m_t=250$GeV, 16\% is found.


As a conclusion, we have given the full leading log QCD
corrections(include
QCD runnings from $m_{top}$ to $M_W$), with whole anomalous dimension
matrix untruncated. Comparison to the previous calculation\cite{Cho},
two points are improved:

(1) Correct errors of anomalous dimensions in ref.\cite{Cho}.

(2) Use untruncated anomalous dimensions in QCD running from
$M_W$ to $m_b$ instead of truncated ones.

In fact, point(1) makes an enhancement while point(2) leads to a
suppression. The total result does not change a lot, e.g. a suppression
of 3\% at $m_{top}=170$GeV comparing ref.\cite{Cho}.

The whole QCD-enhancement of
the  $BR(\overline{B} \rightarrow X_s \gamma)$ makes a factor of 3.9
at $m_t=170$GeV, when $\Lambda_{QCD}=175$MeV.  Using the experimental
branching ratio $BR(\overline{B} \rightarrow X_c e \overline{\nu})
=10.8 \%$, one can find that $BR(\overline{B} \rightarrow X_s \gamma)
\simeq 4 \times 10^{-4}$ at $m_t=174$GeV. It just reaches the upper
limit of present experiment of CLEO. That shows there is very little
space for new physics.

\bigskip

{\bf Acknowledgement}

The authors are grateful to Prof. Xiaoyuan Li for stimulating of this
 work and Prof. Zhaoming Qiu for helpful discussions.


\bigskip

{\bf Figure Captions}

Fig.1 One of the Feynman diagram in calculating Anomalous dimensions,
with the heavy dot denoting high dimension operator.

Fig.2 The photon and gluon magnetic moment operator's coefficient
$C_7(M_W)$(upper) and $C_8(M_W)$(lower) for
different top quark mass. The ones with and without QCD corrections
are indicated by solid and dashed lines respectively. ($\Lambda _{QCD}
=300$MeV is used)

Fig.3 BR($\overline{B} \rightarrow x_s \gamma$) normalized to
	BR($\overline{B} \rightarrow x_c e \overline{\nu}$),
	as function of top quark mass. The upper solid lines indicated
	our results for a full QCD correction.
	Dashed lines correspond to Misiak's results without QCD running
	from $m_{top}$ to $ M_W$.

\end{document}